\documentclass[12pt]{iopart}

\usepackage{amsmath}
\usepackage{amssymb}
\usepackage{xcolor}
\usepackage{graphicx}

\begin{document}

\title[Aggregation, fragmentation, and exchange]{A combined model of aggregation, fragmentation, and exchange processes: insights from analytical calculations
}

\author{Dominic T Robson, Andreas C W Baas, Alessia Annibale}

\address{King's College London, Bush House, North East Wing, 40 Aldwych, London, WC2B 4BG}
\ead{dominic.robson@kcl.ac.uk}
\vspace{10pt}
\begin{indented}
\item[]July 2020
\end{indented}

\begin{abstract}

\noindent{We introduce a mean-field framework for the study of systems of interacting particles sharing a conserved quantity. The work generalises and unites the existing fields of asset-exchange models, often applied to socio-economic systems, and aggregation-fragmentation models, typically used in modelling the dynamics of clusters. An initial model includes only two-body collisions, which is then extended to include many-body collisions and spontaneous fragmentation. We derive self-consistency equations for the steady-state distribution, which can be solved using a population dynamics algorithm, as well as a full solution for the time evolution of the moments, corroborated with numerical simulations.  The generality of the model makes it applicable to many problems and allows for the study of systems exhibiting more complex interactions that those typically considered. The work is relevant to the modelling of barchan dune fields in which interactions between the bedforms and spontaneous fragmentation due to changes in the wind are thought to lead to size-selection.  Our work could also be applied in finding wealth distributions when agents can both combine assets as well as split into multiple subsidiaries.}

\end{abstract}


\section{Introduction}
Many complex systems can be modelled as a collection of particles which interact with one another via a set of rules that transfer some conserved quantity. Such models can yield non-trivial steady-states which often display out-of-equilibrium behaviours \cite{Book}. Exchange models in which assets are transferred between interacting agents have been highly successful in reproducing aspects of socio-economic systems, such as wealth distributions and class formation \cite{ExchangeBoltzmann,ExchangeKinetic,KrapivskyBias,ExchangeReview05,ExchangeGammaDerivations,ExchangeSavings,Inheritance,WealthNetwork2,ClassFormation,ExchangeKrapivsky, BenNaimGelation, Line}. The simplest such models involve unbiased exchanges that can yield Gibbs-like distributions \cite{ExchangeKinetic} or, more generally, produce wealth distributions which can be empirically fit to a gamma distribution \cite{ExchangeGammaDerivations}.  Further complexity can be added by allowing for biased interactions (e.g. favouring the wealthier agent \cite{KrapivskyBias}) and spontaneous splitting of an agent to simulate inheritance \cite{Inheritance}. More recent work has implemented asset-exchange models to study how regulation \cite{Cardoso} and risk taking \cite{Nener} may influence the economy.\\

Another class of models are aggregation-fragmentation models which have been used in a wide variety of applications, ranging from biology to astrophysics \cite{Book,CellDivision,BallisticAggFrag,AggFrag,AggFrag2,SaturnsRings,Prions}.  Such models usually treat agents as polymers consisting of a number of indivisible monomers to simulate processes such as spontaneous fragmentation of a polymer into monomers \cite{AggFrag} (similar to the inheritance model for wealth), and collisions between polymers which can result in constituent monomers joining together or reconfiguring in different groupings \cite{Book}. Analytical expressions for the distributions resulting from aggregation-fragmentation models are possible only for a very small number of interaction kernels \cite{Book}.\\

Many real-world systems evolve through a combination of aggregation, fragmentation, {\it and} exchange processes.  One notable example is fields of barchan sand dunes.  These fast-moving bedforms regularly collide with one another, resulting in aggregation, fragmentation, or exchange of sand \cite{Genois, DuranMF}, while changes in the wind may also trigger spontaneous fragmentation of the dunes \cite{Calving}.  There has been some success in modelling barchan swarms as interacting many-body systems \cite{Genois, DuranMF, Lima, ParteliHerrmann} but a quantitative understanding of how these processes affect size selection remains lacking.  The migration rates of barchans are directly determined by their size and understanding these speeds is key in evaluating the risk the bedforms pose to human activities \cite{Risk1, Risk2}.  An effective model of the size selection process in barchan fields is therefore of great importance.\\

Despite the physical relevance and importance of models combining aggregation, fragmentation, and exchange processes, such models are rare a typically limited to empirical descriptions of numerical results \cite{Inheritance, Genois}.  In this work we present a general model for systems of many clusters interacting through aggregation, fragmentation, and exchange of mass and 
demonstrate that it is possible to derive analytical results for the resulting size distribution in terms of the interactions between clusters.  While motivated by potential applications to barchan dunes, the model we present here is general and should provide a useful theoretical framework to assess the dynamics of many-body systems beyond the field of geomorphology, 
including physical and socio-economic settings.  The remainder of this manuscript is organised as follows:  in section 2 we examine the case of two-body collisions, calculating the steady-state distribution and time-dependent moments and confirming our predictions through numerical simulation.  In section 3 we generalise the model to allow for any processes including spontaneous fragmentation and many-body collisions.  The results are summarised and discussed in section 4. Technical details are provided in the appendices.

\section{Two-body collisions model}

We consider a system of interacting particles (dunes, people, etc.) among which some continuous quantity (mass, wealth, etc.) is shared; we call this quantity volume.  Collisions take place between pairs of particles and result in merging, volume exchange, or the creation of a third particle due to fragmentation.  Volume is conserved in all collisions such that the total volume is constant.  Particle number is not conserved due to merging and fragmentation events.  All pairs of particles are equally likely to collide such that, in a system of $N$ particles, collisions occur at a rate $\alpha_{tot}N(N-1)/2$ where the rate coefficient, $\alpha_{tot} = \alpha_m + \alpha_e + \alpha_f$, is the sum of the rate coefficients for merging, exchange, and fragmentation.\\

\subsection{Output channels}

There are six different forms of outgoing volume: $v_1$ is the result of merging, $v_2$ and $v_3$ are the results of an exchange collision, and $v_4$, $v_5$, and $v_6$ are the outputs of a fragmentation event.  Each of the outputs has the form $v_i = k_i v_{tot}$ where $v_{tot}$ is the total volume of the colliding particles, $v_{tot} = v_a + v_b$, and $k_i \in [0, 1]$.  In the case of merging we have simply $k_1 = 1$ but the situation is more complicated for the other interactions.  The coefficients $k_2$ and $k_3$ are given by

\begin{equation}
    k_2 = \frac{1}{1 + r_e} \qquad \text{and} \qquad k_3 = \frac{r_e}{1 + r_e},
\end{equation}

\noindent{such that $r_e$ is the ratio of the output volumes, which we assume is a stochastic variable drawn from the distribution, $p_e(r_e; v_a, v_b)$, which may depend on the input volumes.  Our choice of the stochastic variable is not unique but was chosen to align with studies of barchan dunes \cite{DuranMF}.  In asset-exchange models it is more common to use the amount of volume exchanged \cite{ClassFormation}.  The different formalisms are easily recoverable from one another by a change of variables in the distribution.  For a fragmentation collision, two such stochastic variables, $r_{f1}$ and $r_{f2}$, are drawn from distributions $p_{f1}(r; v_a, v_b)$ and $p_{f2}(r; v_a, v_b)$. The variable $r_{f1}$ is the ratio between one outgoing volume and the sum of the other two, while $r_{f2}$ is the ratio between those two particles i.e. the outputs are $k_4 v_{tot}$, $k_5 v_{tot}$ and $k_6v_{tot}$ where}

\begin{equation}
    k_4 = \frac{1}{1 + r_{f1}}, \qquad k_5 =\frac{r_{f1}}{1 + r_{f1}}, \qquad \text{and} \qquad k_6 = \frac{r_{f1}r_{f2}}{(1 + r_{f1})(1 + r_{f2})}.
\end{equation}

\noindent{We call the different forms of $v_i \equiv k_i v_{tot}$ the \textit{output channels} of our model and define the \textit{channel probabilities}}

\begin{equation}\label{eq: ChannelRates}
    p_i = \begin{cases}
    \frac{\alpha_m}{\alpha_{out}} \qquad \text{if} \qquad i = 1\\
    \frac{\alpha_e}{\alpha_{out}} \qquad \text{if} \qquad i = 2, 3\\
    \frac{\alpha_f}{\alpha_{out}} \qquad \text{if} \qquad i = 4, 5, 6
    \end{cases}
\end{equation}

\noindent{where $\alpha_{out} \equiv \alpha_m + 2 \alpha_e + 3 \alpha_f$.  The probabilities describe the relative prevalence of each output channel in the population.}\\

\subsection{Master equation}

The distributions of $r_e$, $r_{f1}$, and $r_{f2}$ are the \textit{collision rules} of the model and, in general, depend upon the input volumes $v_a$ and $v_b$.  We, however, consider the case where the collisions are fully random i.e. the distributions are not functions of the colliding volumes, this is typical for asset-exchange models \cite{ExchangeKinetic} and has also been used in a barchan dune model \cite{Genois}.  With this simplification, we can write an expression for the probability distribution of the outputs of a collision.  To do this we perform a sum over the output channels weighted by the channel probabilities, and average over the distributions of the stochastic variables.  The probability, $p_{gain}(v,t)dv$, of a collision creating a particle with volume in the range $[v, v+dv]$ is therefore

\begin{multline}\label{eq: PGainDef}
    p_{gain}(v, t)dv = \sum_{i = 1}^{6} p_i \Bigg\{\int dr_e p_e(r_e) \int dr_{f1}p_{f1}(f_{f1}) \int dr_{f2}p_{f2}(r_{f2})\\
     \int dv_{a} p(v_a, t)\int dv_{b} p(v_b, t)\delta\big((v_a + v_b)k_i(r_e, r_{f1}, r_{f2}) - v\big) \Bigg\}dv,
\end{multline}

\noindent{where $\delta(...)$ is the Dirac $\delta$-function, and $p(v,t)$ is the probability density function (pdf) of volumes in the system at time $t$.  We have assumed that the system is sufficiently large that $v_a$ and $v_b$ can be treated as independently distributed. One can see that \eqref{eq: PGainDef} is simply the expectation of a $\delta$-function}

\begin{equation}\label{eq: PGain}
    p_{gain}(v,t)= \Big\langle \delta\big((v_a + v_b)k_i(r_e, r_{f1}, r_{f2}) - v\big) \Big\rangle_{i, r_e, r_{f1}, r_{f2}, v_a, v_b},
\end{equation}

\noindent{where we have used the shorthand $\langle f(x) \rangle_x \equiv \int dx p(x) f(x)$ for continuous variables or $\langle f_i \rangle_i \equiv \sum_i p_i f_i$ for discrete variables.  We note that we could write $p_{gain}$ in this form solely because of the output channel notation; had we summed over the processes rather than channels, then every term would have featured different numbers of $\delta$-functions equal to the number of outputs for the corresponding process.}\\

The system will lose a particle of volume in the range $[v, v+dv]$ if that particle is involved in a collision.  Since each collision occurs between a pair of particles, the probability of a collision destroying such a particle is approximately $2p(v,t)dv$, where we have again assumed that the system size is sufficiently large.  By combining the gain and loss terms we can now write the master equation describing the average behaviour of the volume frequency density i.e. the number of particles, $N(v,t)dv$, in the range $[v, v+dv]$ at time $t$

\begin{equation}\label{eq: NVT}
    \dot{N}(v,t)dv = \frac{N(t)(N(t) - 1)}{2}\big[\alpha_{out} p_{gain}(v,t) - 2\alpha_{tot}p(v,t)\big]dv,
\end{equation}

\noindent{where the average is over many realisations of the system.  Integrating over the volume, we can get an expression for the time evolution of the average population size}

\begin{equation}\label{eq: DetNT}
    \dot{N}(t) = \frac{N(t)(N(t) - 1)}{2}\big[\alpha_f - \alpha_m\big].
\end{equation}

\noindent{It is important to note here that the population size in individual realisations of the system will fluctuate about this average value and will depend on the exact collisions that occur.  For now, we focus on the deterministic evolution of the average behaviour of the system, described in equations \eqref{eq: NVT} and \eqref{eq: DetNT}, leaving discussion of the time evolution of the fluctuations until section 2.4.  From our expressions for $\dot{N}(v,t)$ and $\dot{N}(t)$ we can write the master equation for the volume pdf}

\begin{equation}\label{eq: PVT}
    \dot{p}(v,t) = \frac{\dot{N}(v,t)}{N(t)} - \frac{\dot{N}(t)}{N(t)} p(v,t) =(N(t) - 1)\frac{\alpha_{out}}{2}\big[p_{gain}(v,t) - p(v,t)\big].
\end{equation}

\noindent{The steady-state volume pdf is obtained by setting the right hand side (RHS) of equation \eqref{eq: PVT} to zero, however one must be careful to ensure that such a state is possible.  Since volume is conserved in all collisions, the mean volume is governed solely by the behaviour of $N^{-1}(t)$, hence it is only constant if the population size is constant.  For $\alpha_m \neq \alpha_f$, the only fixed point of equation \eqref{eq: DetNT} is $N(t) = 1$, corresponding to a $\delta$-function volume pdf located at the total initial volume.  On the other hand, for $\alpha_f = \alpha_m$, equation \eqref{eq: DetNT} implies that the average population size will remain constant at the initial size.  In this case, equation (8) allows for a non-trivial steady-state for the volume pdf, obeying}

\begin{equation}\label{eq: Steady}
    p_s(v) = p^{(s)}_{gain}(v) = \Big\langle \delta((v_a + v_b)k_i - v) \Big\rangle^{(s)}_{i, r_e, r_{f1}, r_{f2}, v_a, v_b},
\end{equation}

\noindent{where the subscript $s$ and superscript $(s)$ indicate that this is in the steady-state.  The distributions of the input volumes, $v_a$ and $v_b$, which are averaged over on the RHS of equation \eqref{eq: Steady}, are the same as the distribution appearing on the LHS.  Therefore, equation \eqref{eq: Steady} is a self-consistency equation which can be solved using an iterative algorithm.  We were able to solve this equation using a similar approach to the population dynamics algorithm described in \cite{PopDyn1} and \cite{PopDyn2} which has been developed in statistical physics to solve self-consistency equations for distributions, as in equation \eqref{eq: Steady}.}\\

We are able to obtain the non-trivial steady-state above only because the average population size remains constant when $\alpha_m = \alpha_f$.  However, $N = 1$ is an absorbing state of the system since collisions cannot occur below $N = 2$, hence, even with equal rates of merging and fragmentation, the system size of an individual realisation will eventually converge to this absorbing state via stochastic fluctuations.  Therefore, our assumption of constant population will hold for individual realisations only for timescales which are short compared to the initial population size.  In applications to physical systems such as barchan swarms, the timescales over which the steady-state will persist will be very large such that, for practical purposes, one can treat the system as remaining in this steady-state.\\  

In addition to the exact solution for the steady-state volume pdf, we are able to calculate the moments of the time-dependent distribution.  This can be done for any values of $\alpha_m, \ \alpha_f$, and $\alpha_e$ however the most interesting case is when $\alpha_m = \alpha_f$.  Since the population size is, on average, constant, the mean volume, $\langle v \rangle$ is also constant.  From equation \eqref{eq: PVT}, the higher integer moments evolve as 

\begin{equation}\label{eq: EOMMoms}
    \frac{d\langle v^\ell \rangle}{dt} = (N_0 - 1)\alpha_{tot}\left[K_\ell\sum_{j = 1}^{\ell-1}{\ell \choose j}\langle v^j\rangle \langle v^{\ell - j}\rangle + (2K_\ell - 1)\langle v^\ell \rangle\right],
\end{equation}

\noindent{where $N_0$ is the population size and}

\begin{equation}
K_\ell \equiv \langle k_i^\ell (r_e, r_{f1}, r_{f2})\rangle_{i, r_e, r_{f1}, r_{f2}}.
\end{equation}

\noindent{Equation \eqref{eq: EOMMoms} has the form $\dot{x} = ax + b(t)$ and so can be easily solved to give}

 \begin{equation}\label{eq: MomT}
     \langle v^\ell \rangle(t) = \left[\langle v^\ell \rangle_0 - \frac{\beta_\ell K_\ell}{1 - 2K_\ell}\sum_{j = 1}^{\ell-1}{\ell \choose j}\int_0^t dt'\langle v^j\rangle(t')\langle v^{\ell - j}\rangle(t')e^{-\beta_\ell t'}\right]e^{\beta_\ell t},
 \end{equation}

\noindent{where}

\begin{equation}
    \beta_\ell = (N_0 - 1)(2K_\ell - 1)\alpha_{tot}.
\end{equation}

\noindent{As $t$ increases, equation \eqref{eq: MomT} converges to a steady-state which can also be calculated directly from equation \eqref{eq: EOMMoms}}
 
 \begin{equation}\label{eq: SSMoms}
     \langle v^\ell \rangle_s = \frac{K_\ell}{1 - 2K_\ell}\sum_{j = 1}^{\ell-1}{\ell \choose j}\langle v^j\rangle_s\langle v^{\ell - j}\rangle_s,  
 \end{equation}
 
 \noindent{where we use the subscript $s$ to denote that this is in the steady-state.  One can show that $2K_\ell < 1 \ \forall \ \ell > 1$ in the case of balanced merging and fragmentation such that the moments are always positive.  Equations \eqref{eq: MomT} and \eqref{eq: SSMoms} can be solved sequentially up to any desired moment.  An important implication of equation \eqref{eq: SSMoms} is that $\ell$-th moment of the steady-state is proportional to the mean to the $\ell$-th power, $\langle v^\ell \rangle_s \propto \langle v \rangle_s^\ell$, with the proportionality constant depending upon the collision rule.  Proportionality was previously reported in a numerical study \cite{DuranMF} where a partially-deterministic collision rule was used, this suggests that an expression similar to equation \eqref{eq: SSMoms} may still hold even when collisions depend on the input volumes.}  \\

\subsection{Numerical results: population dynamics and Gillespie algorithm}

To verify the results derived above we simulated the system using a Gillespie algorithm \cite{Gillespie}.  We tested many different initial conditions, rate coefficients, and stochastic variable distributions and found that the behaviour of the system, averaged over a number of repeated simulations, agreed well with the theory.  In figure 1 we show the time evolution of the scaled second and third moments in the Gillespie simulations and the theoretically predicted values given by equation \eqref{eq: MomT}.\\  

\begin{figure}[h]
    \centering
    \includegraphics[scale = 0.8]{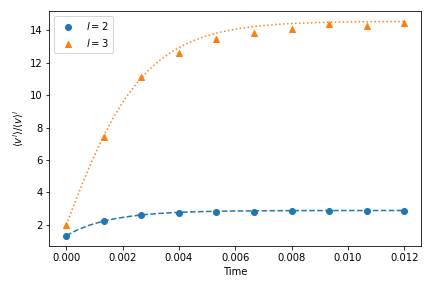}
    \caption{The time evolution of the relative second and third moments $\langle v^\ell \rangle/\langle v \rangle^\ell$ with $\ell = 2, \ 3$ from Gillespie simulations (points) compared to the prediction (lines) from equation (12).  The data are the mean of 250 simulations, with $2.5 \times 10^3$ initial particles with uniformly distributed volume between 0 and 1.  The rate coefficients were set to be $\alpha_e = 0.25$ and $\alpha_m = \alpha_f = 0.375$.  The distributions $p_e$, $p_{f1}$, and $p_{f2}$ were uniform in the range $[0,1]$.   Time was defined in the Gillespie algorithm such that the average time between collisions was $2/N(N-1)$}
    \label{fig: figure1}
\end{figure}

\begin{figure}[h]
    \centering
    \includegraphics[scale = 0.9]{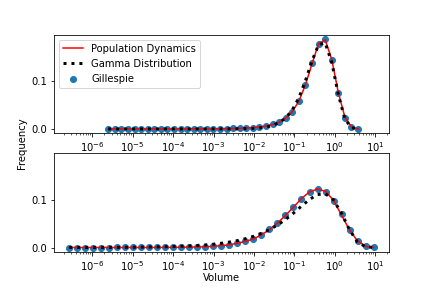}
    \caption{Average histograms of the final states of 250 Gillespie simulations compared with steady-state solution solved using population dynamics and gamma distributions estimated from the theoretically predicted moments.  The top figure is the case of exchange only with $\alpha_e = 1$, $\alpha_m = \alpha_f = 0$ while the bottom figure included all three processes with $\alpha_e = 0.25$, $\alpha_m = \alpha_f = 0.375$.  Initially there were $2.5 \times 10^3$ particles with volumes uniformly distributed in $[0,1]$.  The distributions $p_e$, $p_{f1}$, and $p_{f2}$ were uniform in the range $[0,1]$.  The bins were spaced equally along the logarithmic volume axis.  Both $x$ and $y$-axes have been set to the same scale so that the figures can be compared.}
    \label{fig: Figure2}
\end{figure}

In Figure \ref{fig: Figure2} we plot the steady-state distribution obtained from the Gillespie simulations for different values of the model rates, and the analytical expression given in equation \eqref{eq: Steady} solved using the population dynamics algorithm \cite{PopDyn1, PopDyn2} (see Appendix A for details of the algorithm).  The steady-states were computed by running many simulations and taking the average over all of the final states.  Results from simulations were found to be in excellent agreement with the theory.  When only exchange collisions were included the distribution was well described by a gamma distribution whose parameters could be predicted from the moments (see Appendix B) given by equation \eqref{eq: SSMoms}.  Gamma distributions are common results in asset-exchange models \cite{ExchangeGammaDerivations} however the fit was not so strong when the other processes were included.  With our choices of collision rules, merging and fragmentation were found to lead to a broader distribution than when only exchange collisions occurred.

\subsection{Beyond the deterministic approximation}

So far we have assumed that the system evolved deterministically according to equations \eqref{eq: NVT} and \eqref{eq: DetNT}, here we briefly discuss some properties of the system beyond this approximation.\\

The probability $p_N(t)$ that the system contains $N$ particles at time $t$ evolves according to the master equation 

\begin{equation}\label{eq: MEPNT}
    \dot{p}_N(t) = \frac{N(N+1)\alpha_m}{2}p_{N+1} + \frac{(N-1)(N-2)\alpha_f}{2}p_{N-1} - \frac{N(N-1)(\alpha_f + \alpha_m)}{2}p_N.
\end{equation}

\noindent{Despite the non-linear dependence on $N$ we were able to obtain an analytical solution for the distribution $p(x,t)$ with $x \equiv N/N_0$, within a Kramers-Moyal (KM) approximation scheme \cite{Kramers, Moyal}.  As before, we are most interested in balanced rates of merging and fragmentation $\alpha_f = \alpha_m = \alpha$.  Since transitions between states are $N \rightarrow N \pm 1$, transitions in $x$ are of order $N_0^{-1} \ll 1$, the KM expansion then gives}

\begin{equation}
    \partial_t p(x,t) = \frac{\alpha}{2}\partial_x^2[x^2p(x,t)] + O(N_0^{-1}),
\end{equation}

\noindent{where we have assumed that $N \approx N_0 \gg 1$ such that $x \approx 1$ which is valid at early times.  Neglecting $O(N_0^{-1})$ terms we Laplace transform the time variable to give the ODE}

\begin{equation}\label{eq: LaplaceODE}
    sP(x,s) - \delta(x-1) = \frac{\alpha}{2}\partial_x^2[x^2P(x,s)],
\end{equation}

\noindent{where $P(x,s) \equiv \mathcal{L}_t[p(x,t)](x,s)$ is the Laplace transformed distribution and we have used $p(x, 0) = \delta(x-1)$.  We can solve equation \eqref{eq: LaplaceODE} using the boundary condition}

\begin{equation}\label{eq: BoundaryCondition}
    p(x, 0) = 0 \ \forall \ x \neq 1 \ \implies \ \lim_{s \rightarrow \infty} s P(x,s) = 0 \ \forall \ x \neq 1,
\end{equation}

\noindent{which yields the solution (see Appendix C for details)}

\begin{equation}\label{eq: LapSol}
    P(x,s) = \frac{2}{\alpha(\lambda_+(s) - \lambda_-(s))}\big[(1 - \theta(x-1))x^{\lambda_+(s)} + \theta(x-1)x^{\lambda_-(s)}\big],
\end{equation}

\noindent{where  $\theta(x)$ is the Heaviside step-function and}

\begin{equation}\label{eq: LambdaPM}
    \lambda_{\pm}(s) = -\frac{3}{2} \pm \frac{1}{2}\sqrt{1 + \frac{8s}{\alpha}}.
\end{equation}

\noindent{One can check that equation \eqref{eq: LapSol} corresponds to the Laplace transformed probability distribution since integrating gives $\int_0^\infty dx P(x,s) = s^{-1} = \mathcal{L}[1](x,t)$.  Finally, it is possible to find the inverse Laplace transform of $P(x,s)$ to recover $p(x,t)$}

\begin{align}\label{eq: XDistribution}
    p(1,t) &= \frac{1}{\sqrt{2 \pi \alpha t}} \exp\left[-\frac{\alpha t}{8}\right],\\
    p(x \neq 1,t) &= \frac{1}{2\pi\alpha}\frac{\mid \log x \mid}{\sqrt{x^3}}\exp\left[-\frac{\alpha t}{8}\right]\int_0^t d\tau \frac{1}{\sqrt{(t - \tau)\tau^3}}\exp\left[-\frac{\log^2 x}{2 \alpha \tau} \right].
\end{align}

\noindent{We stress here that this derivation has assumed $p(x,t)$ is strongly peaked close to $x = 1$ however equation (22) shows that the distribution is widening with time so that this approximation will eventually break down.}\\

We simulated the dynamics described by equation \eqref{eq: MEPNT} using a Gillespie algorithm.  In figure 3 we show a snapshot of the distribution $p(x,t)$ for different values of $\alpha_m = \alpha_f = \alpha$ which is equivalent to looking at snapshots of a simulation at different times.  One can see that the simulated results are well described by the theory of equations (21) and (22) and that the distribution is wider for larger values of $\alpha$ which is equivalent to width increasing with time.\\

\begin{figure}[h]
    \centering
    \includegraphics[scale = 0.8]{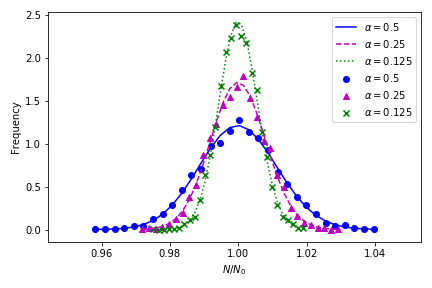}
    \caption{Histograms showing the distribution of $x = N/N_0$ at simulation time $2.5 \times 10^{-4}$ for three different values of $\alpha_m = \alpha_f = \alpha$.  The histograms come from 5000 repeats of simulations initiated with $10^4$ particles.  The lines shown are prediction of equations (21) and (22).}
    \label{fig: Figure3}
\end{figure}

Instead of performing a KM expansion we can also use equation \eqref{eq: MEPNT} to solve for fluctuations in the inverse population size $N^{-1}$ which is equivalent to fluctuations in the mean particle volume since total volume is conserved.  Again we assume that the distribution $p_N$ is sharply peaked at $N_0 \gg 1$.  We can then use equation \eqref{eq: MEPNT} to find that

\begin{equation}\label{eq: EOMInvN}
    \frac{d \langle N^{-1} \rangle}{dt} = \alpha\big[\langle N^{-1}\rangle - \langle N^{-2}\rangle\big] \quad \text{and} \quad
    \frac{d \langle N^{-2} \rangle}{dt} = 3\alpha\langle N^{-2}\rangle,
\end{equation}

\noindent{where we have neglected any $\mathcal{O}(N^{-3})$ terms (see Appendix D for details).  We can solve this pair of coupled ODEs to give}

\begin{equation}\label{eq: Inv1and2}
    \langle N^{-1}\rangle(t) = \Big[\frac{1}{N_0}+ \frac{1}{2N_0^2}(1 - e^{2\alpha t})\Big]e^{\alpha t}\quad \text{and} \quad
    \langle N^{-2} \rangle(t) = \frac{1}{N_0^2}e^{3 \alpha t}.
\end{equation}

\noindent{Defining the fluctuation in inverse population size, $\xi(t)$, as the average difference between the inverse population size and $N_0^{-1}$ we find}

\begin{equation}\label{eq: XiT}
    \xi(t) \equiv \sqrt{\left\langle(N^{-1}(t) - N_0^{-1})^2\right\rangle} = \frac{1}{N_0}\sqrt{1 - 2e^{\alpha t} + e^{3\alpha t}} \approx \frac{\sqrt{\alpha t}}{N_0},
\end{equation}

\noindent{Again, we see that the fluctuations about the $N^{-1} = N_0^{-1}$ steady state are increasing with time.  This result was checked against our simulations and found to be in good agreement as shown in figure 4.}\\

\begin{figure}
    \centering
    \includegraphics[scale = 0.8]{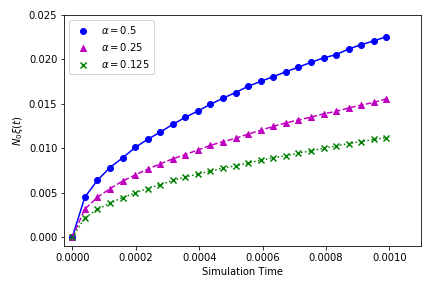}
    \caption{Fluctuations in the inverse population size from Gillespie simulations (points) compared to the prediction of equation (24) (lines).}
    \label{fig: Figure4}
\end{figure}


\section{General processes model}

We now extend our formalism to allow for any $n$-input $m$-output processes, including spontaneous fragmentation which plays an important role in many systems \cite{Inheritance, AggFrag, Calving} and many-body collisions which are also highly important, for instance in barchan dune fields \cite{ParteliAsymmetry}.\\

\subsection{Master equation}

To generalise the rates of the processes we introduce rate coefficients $\alpha_{nm}$ such that the rate of $n \rightarrow m$ processes in a system of $N$ particles is

\begin{equation}
    \text{Rate}(n \rightarrow m) = \alpha_{nm}{N \choose n}
\end{equation}

\noindent{Collisions conserve volume meaning that $n \neq 0$ and $m \neq 0$ but all other processes are permitted.}\\

We explained in the original model that output channel notation is necessary to write the steady-state in terms of the expectation of a $\delta$-function because it allowed us to take averages over the channels rather than the processes.  We can do this in the general model if we write everything as a function of the channel number.  We first need to establish how many channels we need, this will be equal to the total number of outputs of all processes.  In our original model we had a one-output process (merging), a two-output process (exchange), and a three-output process (fragmentation), hence there were six output channels, each corresponding to a single output of a process.  By specifying the channel number, $i$, we can determine the number of inputs, $n_i$, and the number of outputs, $m_i$, that lead to that channel.  For instance, in our original model, channel 1 was the sole output of the $2 \rightarrow 1$ merging collision hence $n_1 = 2$ and $m_1 = 1$, while channel 2 was one of the outputs of the $2 \rightarrow 2$ exchange collision so $n_2 = 2$ and $m_2 = 2$.  One will notice that an $n \rightarrow m$ process contributes $m$ channels to the model and so there are $m$ values of $i$ for which $n_i = n$ and $m_i = m$ i.e.  

\begin{equation}\label{eq: ChannelSumProperty}
    \sum_j \delta_{n, n_j} \delta_{m, m_j} = m,
\end{equation}

\noindent{where $\delta_{i,j}$ is the Kronecker delta and the sum runs over all of the channels.  One can easily check that this relation holds for our original model where $n_{1,2,3,4,5,6} = 2$, $m_1 = 1$, $m_{2,3} = 2$ and $m_{4,5,6} = 3$.}\\

In the original model the output channel volumes had the form $k_i v_{tot}$ where $v_{tot} = v_a + v_b$ was the same for all channels, since every channel had $n_i = 2$.  In the generalised model the number of inputs is not the same for every channel and so the output channel volumes are $k_i v_{tot}^{(i)}$ with

\begin{equation}
    v_{tot}^{(i)} \equiv \sum_{j = 1}^{n_i}v_{a_j}^{(i)},
\end{equation}

\noindent{where the $v_{a_j}^{(i)}$ are the $n_i$ inputs of the $n_i \rightarrow m_i$ process.  As before, $k_i$ depends on stochastic variables.  We group these variables into a stochastic vector $\vec{r}$ e.g. in our first model $\vec{r} = (r_e, r_{f1}, r_{f2})$.  We again assume that the distribution, $p_r(\vec{r})$, is independent of the input volumes i.e. fully random collisions.}\\

Using shorthand $\alpha_i \equiv \alpha_{n_i m_i}$, channel $i$ occurs at a rate $\alpha_i {N(t)\choose n_i}$.  The channel probability of channel $i$ is given by this rate divided by the sum of the rates of all channels, that is 

\begin{equation}
    p_i(t) = \frac{\alpha_i{N(t) \choose n_i}}{\sum_j \alpha_j {N(t) \choose n_j}},
\end{equation}

\noindent{where the sum over $j$ runs over all possible channels.  One can check by inserting the relevant terms from the original model that the channel probabilities agree with those we had written in equation (3).}\\

The derivation of the master equations follows in a very similar manner to the previous section.  We were able to find the generalisations of equations \eqref{eq: DetNT} and \eqref{eq: PVT}

\begin{align}\label{eq: GeneralModel}
    \dot{N}(t) &= \sum_i{\frac{\alpha_{i}}{m_i}{N(t) \choose n_i}(m_i - n_i)},\\
    \dot{p}(v,t) &= \frac{1}{N(t)}\big[p_{gain}(v,t) - p(v,t)\big]\sum_j\alpha_j {N(t) \choose n_j}
\end{align}

\noindent{where, again, the sums run over all channels and}

\begin{equation}\label{eq: PVTGeneralModel}
    p_{gain}(v,t) = \Big\langle \delta\big(k_i\left(\vec{r}\right)v_{tot}^{(i)} - v\big) \Big\rangle_{i, \vec{r}, v_{a_1}^{(i)},...,v_{a_{n_i}}^{(i)}}(t).
\end{equation}

\noindent{Again, the form of $p_{gain}$ as the expectation of a $\delta$-function is possible because we average over the output channels rather than the processes themselves.}\\

A steady-state solution,  $p(v,t) = p_s(v)$, can only occur if there is a value of $N = N_s$ which is the solution to $\dot{N} = 0$.  Assuming that the steady-state exists, we again obtain a self-consistency equation for $p_s(v)$ in the form of the expectation of a $\delta$-function

\begin{equation}\label{eq: PsteadyGeneral}
    p_s(v) = p_{gain}^{(s)}(v) = \Big\langle \delta\big(k_i\left(\vec{r}\right)v_{tot}^{(i)} - v\big) \Big\rangle_{i, \vec{r}, v_{a_1}^{(i)},...,v_{a_{n_i}}^{(i)}}^{(s)}, 
\end{equation}

\noindent{that can be solved, as earlier, with a population dynamics algorithm.  The mean volume of the system is determined as $\langle v \rangle_s \propto N_s^{-1}$ where the constant of proportionality is simply the total volume of all particles, which is constant.  We can also obtain expressions for the higher moments of the steady-state distribution}

\begin{equation}\label{eq: SteadyMomsGen}
    \langle v^\ell \rangle_s = \frac{1}{Z_\ell}\sum_i p_i \langle k_i^\ell \rangle_{\vec{r}} \sum_{\substack{j_1 + ... + j_{n_i} = \ell\\ j_1, ..., j_{n_i} \neq \ell}}{\ell \choose j_1...j_{n_i}}\langle v^{j_1}\rangle_s...\langle v^{j_{n_i}}\rangle_s ,
\end{equation}

\noindent{where ${\ell \choose j_1...j_{n_i}}$ are the multinomial coefficients and} 

\begin{equation}
    Z_\ell = 1 - \sum_i p_i n_i \langle k_i^\ell \rangle_{\vec{r}}.
\end{equation}

\subsection{Numerical results}

\begin{figure}[h]
    \centering
    \includegraphics[scale = 0.95]{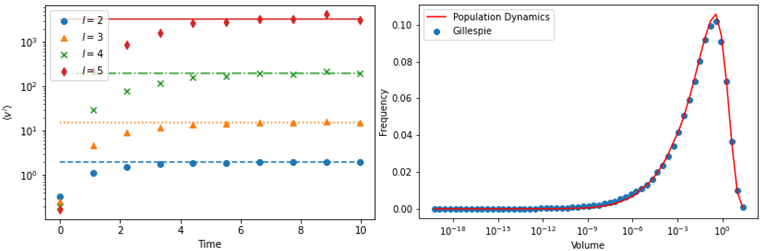}
    \caption{Average results from 150 Gillespie simulations with rate coefficients $\alpha_s = 1.0$, $\alpha_e = 1.0 \time 10^{-5}$, $\alpha_t = 1.3 \times 10^{-5}$, and $\alpha_m = \alpha_f = 0$ leading to a steady-state population size $N_s = 1500$.  Systems were initialised at the steady size, with volumes distributed uniformly between zero and unity.  All stochastic variables were uniform between zero and unity.  Left:  Integer moments $\langle v^\ell \rangle_s$ with $\ell = 2,3,4,5$ from the simulation and the values predicted from equation (34), shown as lines.  Right: Average histogram of the final state of the Gillespie simulations compared to the prediction of equation (33) solved using population dynamics.  The bins were spaced uniformly in logarithmic space.}
    \label{fig: Figure5}
\end{figure}

In this section we verify the theoretical results derived for the general model by means of Gillespie simulations.  To demonstrate the sorts of new processes that can be studied using our general model, we added $1 \rightarrow 2$ spontaneous fragmentation and $3 \rightarrow 1$ three-body merging.  To match some of the notation of the previous section we labelled the rate coefficients $\alpha_{21} = \alpha_m$, $\alpha_{22} = \alpha_e$ and $\alpha_{23} = \alpha_f$, $\alpha_{12} =  \alpha_s$ and $\alpha_{31} = \alpha_t$ where $s$ stands for spontaneous fragmentation and $t$ for three-body merging.  We simulated many different systems for which a steady-state was reached, verifying the Gillespie simulations were in agreement with the theory, solved using population dynamics (see figure \ref{fig: Figure5}).  We also confirmed that the moments converged to those predicted by equation \eqref{eq: SteadyMomsGen} as shown in figure \ref{fig: Figure5}.

\section{Discussion and conclusion}

In this work we have introduced a general framework for the study of many-body systems of particles in which random $n$-input, $m$-output processes occur.  Such processes have been previously studied in asset-exchange and aggregation-fragmentation models used in a broad range of applications including socio-economic systems, barchans dunes, and systems of interacting polymers.  Our work represents a novel generalisation of asset-exchange and aggregation-fragmentation models, combining the two into a single framework in which analytical results are tractable.\\

We have shown that, in the case of random interactions, it is possible to derive analytical expressions for the steady-state moments as well as a self-consistency equation for the steady-state distribution which we were able to solve using an iterative algorithm borrowed from statistical mechanics.  In the case of two-body interactions we have also derived the full time evolution of the moments and we have analysed fluctuations in the population size showing that the fluctuations are time dependent.  All theoretical results have been verified by numerical simulations.\\

Since the results of our theoretical derivations directly relate the steady-state distribution to the choice of interaction rules, our model raises the possibility of inferring properties of interactions from measurement of the steady-state distribution.  In future works we will demonstrate application of our model to the study of barchan swarms, using the predictions of this model to infer the nature of dune interactions in such systems.\\  

The work we have presented here is limited to random collisions, these have had great success in modelling wealth distributions using asset-exchange models and replicating simulated dune-field dynamics.  Nevertheless, allowing for deterministic collisions may well be an interesting avenue to explore.  Further work could also explore the nature of fluctuations in the volume distribution itself, rather than just the population size.  One may also imagine allowing for the annihilation or creation of particles so that total volume is not conserved.  We feel that the potentially wide applicability of the scheme we have introduced warrants further study along such lines.\\

\ack 

DTR is supported by the EPSRC Centre for Doctoral Training in Cross-Disciplinary Approaches to Non-Equilibrium Systems (CANES EP/L015854/1).  We would like to thank an anonymous reviewer for their useful suggestions.

\appendix

\section{Population dynamics algorithm}

Each iteration of the population dynamics algorithm consisted of the following steps:

\begin{enumerate}
    \item Randomly select one output channel $i$ based on the channel probabilities $\{p_i\}$
    \item Randomly select $n_i$ volumes from the current distribution and sum to calculate $v_{tot}^{(i)}$
    \item Draw any random variables $\vec{r}$ relevant to channel $i$ from their respective distributions
    \item Calculate the output volume $v_i = k_i(\vec{r})v_{tot}^{(i)}$
    \item Select a particle (it can be one that was already selected) and set its volume to $v_i$
\end{enumerate}

The algorithm was run for long enough for the second moment to have approximately converged and the average of many runs of the algorithm was taken as the solution for the steady-state. 

\section{Estimating gamma distribution parameters}

For exchange-only processes a gamma distribution was shown to be a good approximation of the steady-state (figure \ref{fig: Figure2}).  The gamma distribution shown had pdf 

\begin{equation}
    p_{\gamma}(v) = \frac{\beta^\alpha}{\Gamma(\alpha)}\exp[-\beta v]v^{\alpha - 1},
\end{equation}

\noindent{where we estimated the parameters $\alpha$ and $\beta$ as}

\begin{equation}
    \alpha = \langle v \rangle_s \beta = \frac{\langle v \rangle_s^2}{\langle v^2 \rangle_s - \langle v \rangle_s^2} = \frac{1 - 2K_2}{4K_2 - 1},
\end{equation}

\noindent{where the final equality is derived directly from equation \eqref{eq: SSMoms}.  Since the mean volume is a constant and the initial volume distribution was uniform in the range $[0,1]$ we took $\langle v \rangle_s = 0.5$.}

\section{Kramers-Moyal differential equation}

The general solutions of the differential equation \eqref{eq: LaplaceODE} have the form

\begin{equation}\label{eq: GeneralPXS}
    P(x,s) = \left(A_{+}(s) - \frac{2\theta(x-1)}{\alpha(\lambda_+(s) - \lambda_-(s))}\right)x^{\lambda_+(s)} +\left(A_{-}(s) + \frac{2\theta(x-1)}{\alpha(\lambda_+(s) - \lambda_-(s))}\right)x^{\lambda_-(s)}, 
\end{equation}

\noindent{where $\lambda_{\pm}(s)$ are defined in equation \eqref{eq: LambdaPM} and $A_{\pm}(s)$ are boundary terms.  The boundary condition, \eqref{eq: BoundaryCondition}, can be used to find these boundary terms for large values of $s$ which is sufficient since large $s$ will be equivalent to early times when we invert the Laplace transform.  The boundary condition holds for all $x \neq 1$ so we consider first $x < 1$ in which case we have}

\begin{equation}
    P(x < 1, s) = A_+ (s) x^{\lambda_+(s)} + A_-(s) x^{\lambda_-(s)}.
\end{equation}

\noindent{We require that this is zero in the limit that $s \rightarrow \infty$ for which $\lambda_\pm (s) \rightarrow \pm \infty$.  Since $x$ is positive and less than unity we therefore have that $x^{\lambda_+} \rightarrow 0$ and $x^{\lambda_-} \rightarrow \infty$ so we find that for large $s$}

\begin{equation}
    P(x < 1,s \rightarrow \infty) = 0 \quad \implies \quad A_-(s) = 0.
\end{equation}

\noindent{To find $A_+(s)$ we look at the case $x > 1$ for which we now have}

\begin{equation}\label{eq: X>1}
    P(x > 1, s \rightarrow \infty) = \left(A_{+}(s) - \frac{2}{\alpha(\lambda_+(s) - \lambda_-(s))}\right)x^{\lambda_+(s)} = 0.
\end{equation}

\noindent{Again we have that $\lambda_+ \rightarrow \infty$ so now $x^\lambda_+ \rightarrow \infty$, for $x > 1$, hence, in order to satisfy equation \eqref{eq: X>1}, we must set}

\begin{equation}
    A_+(s) = \frac{2}{\alpha(\lambda_+(s) - \lambda_-(s))},
\end{equation}

\noindent{for large $s$.  Inserting our values for $A_\pm(s)$ into equation \eqref{eq: GeneralPXS} yields the result we presented in the text, equation \eqref{eq: LapSol}.}\\

The final stage was to invert the Laplace transform in order to find $p(x,t)$.  We were able to find the result using known solutions for the inverse Laplace transforms of standard functions.  

\section{Inverse population size fluctuations}

Starting from equation \eqref{eq: MEPNT} we can write the equation of motion of $\langle N^{-1} \rangle$

\begin{equation}
    \frac{d \langle N^{-1} \rangle}{dt} = \frac{\alpha}{2}\sum_{N=1}^\infty\big[ (N+1)p_{N+1} + (N - 3 + 2N^{-1})p_{N-1} - 2(N-1)p_N\big].
\end{equation}

\noindent{Now we assume that the distribution is strongly peaked close to the initial population size $N_0$ i.e. $p_N = 0$ except for $N \approx N_0$.  This assumption means we are free to shift terms with $p_{N+1}$ such that $N + 1 \rightarrow N$ and the $p_{N-1}$ terms such that $N-1 \rightarrow N$ yielding}

\begin{equation}
    \frac{d \langle N^{-1} \rangle}{dt} = \alpha \sum_{N = 1}^\infty p_N \frac{1}{N + 1}.
\end{equation}

\noindent{Now since we only have finite probabilities close to $N_0$ and we are considering very large populations, $N_0 \gg 1$, we approximate the evolution by Taylor expanding, truncating at $\mathcal{O}(N^{-2})$}

\begin{equation}
    \frac{d \langle N^{-1} \rangle}{dt} = \alpha \sum_{N = 1}^\infty p_N \frac{1}{N}\left(1 - \frac{1}{N} + \frac{1}{N^2} + ...\right) \approx \alpha\big[\langle N^{-1}\rangle - \langle N^{-2}\rangle\big].
\end{equation}

\noindent{The equation of motion for $\langle N^{-2} \rangle$ was derived in the same way to give the results of equation \eqref{eq: EOMInvN}.}\\

\end{document}